\documentclass{article}
\usepackage{spconf,amsmath,graphicx}
\usepackage{multirow}
\usepackage{cite}
\usepackage{threeparttable}
\usepackage{float}
\usepackage{bm}

\title{BOUNDARY AND CONTEXT AWARE TRAINING FOR CIF-BASED NON-AUTOREGRESSIVE END-TO-END ASR}
\name{
\begin{tabular}{c}
    \it Fan Yu$^1$, Haoneng Luo$^1$, Pengcheng Guo$^1$, Yuhao Liang$^1$, Zhuoyuan Yao$^1$, \\
    \it Lei Xie$^{1*}$\thanks{*Lei Xie is the corresponding author, lxie@nwpu.edu.cn}, Yingying Gao$^2$, Leijing Hou$^2$, Shilei Zhang$^2$ \\
\end{tabular}
}

\address{
  $^1$Audio, Speech and Language Processing Group (ASLP@NPU), School of Computer Science,\\ Northwestern Polytechnical University, Xi’an, China \\
  $^2$China Mobile Research Institute}

\begin{document}
%
\maketitle
\begin{abstract}
Continuous integrate-and-fire (CIF) based models, which use a soft and monotonic alignment mechanism, have been well applied in non-autoregressive (NAR) speech recognition with competitive performance compared with other NAR methods. However, such an alignment learning strategy may suffer from an erroneous acoustic boundary estimation, severely hindering the convergence speed as well as the system performance. In this paper, we propose a boundary and context aware training approach for CIF based NAR models. Firstly, the connectionist temporal classification (CTC) spike information is utilized to guide the learning of acoustic boundaries in the CIF. Besides, an additional contextual decoder is introduced behind the CIF decoder, aiming to capture the linguistic dependencies within a sentence. Finally, we adopt a recently proposed Conformer architecture to improve the capacity of acoustic modeling. Experiments on the open-source Mandarin AISHELL-1 corpus show that the proposed method achieves a comparable character error rates (CERs) of 4.9\% with only 1/24 latency compared with a state-of-the-art autoregressive (AR) Conformer model. Futhermore, when evaluating on an internal 7500 hours Mandarin corpus, our model still outperforms other NAR methods and even reaches the AR Conformer model on a challenging real-world noisy test set.

\end{abstract}
\section{Introduction}
\label{sec:intro}
End-to-end (E2E) models have achieved great success in automatic speech recognition (ASR) due to their effectiveness in sequence-to-sequence modeling~\cite{chorowski2015attention,chan2016listen,kim2017joint,chiu2018state,vaswani2017attention,dong2018speech}. Compared with the traditional hybrid systems~\cite{peddinti2015time,povey2016purely}, E2E models can not only simplify the model training process but also achieve competitive or even better recognition accuracy. However, most state-of-the-art E2E models follow an autoregressive (AR) fashion and recursively generate an output token based on previously generated tokens and the input sequence. Thus, it will take at least $L$ iterations to generate an $L$-length output sequence, resulting in a complex computation and a large latency with the increment of sequence length. In contrast, the non-autoregressive (NAR) models make a conditional independence assumption among the output tokens and no longer rely on the temporal relationship from left to right, which can generate an $L$-length sequence in parallel with a constant $K (<< L)$ iterations.

NAR models are originally proposed in neural machine translation (NMT) tasks and have been well applied in ASR recently. The major difficulties of NAR models consist of the following two aspects: the accurate length prediction of a target sequence and the parallel inference of decoder. Introducing a length prediction network behind the encoder~\cite{lee2018deterministic,gu2017non,ma2019flowseq} or estimating an empirical target length according to the source sequence~\cite{bai2020listen} are two typical length prediction approaches. However, both methods will bring redundant computations and can not guarantee the accuracy of the predicted lengths, especially in intricate real-world scenarios. Since connectionist temporal classification (CTC)~\cite{graves2006connectionist} is good at learning frame-wise latent alignments between the input speech and output tokens, more and more studies are trying to get rid of the cumbersome length estimation and focus on incorporating the CTC into NAR models. In~\cite{shi2020sequence, guo2020multi}, conditional chain based methods were proposed for NAR multi-speaker ASR, which inferred the output of each speaker one-by-one using a NAR CTC model. With this design, the total inference steps will be restricted to the number of mixed speakers. In~\cite{tian2020spike}, Tian \textit{et al.} proposed a spike-triggered based NAR method, which used the encoded states corresponding to the CTC spikes as the decoder input. Although the decoder can easily perform parallel computation with this design, the length mismatch between the CTC spikes and target sequence may lead to a computation problem of cross-entropy (CE) loss. Inspired by the masked language modeling, Mask-CTC~\cite{higuchi2020mask,higuchi2020improved} initialized the input target sequence with masked ground-truth during the training and masked token-level CTC outputs during the inference, respectively. The idea of Mask-CTC was to infer the masked tokens by the decoder and iteratively refine them. In addition to Mask-CTC, Imputer~\cite{chan2020imputer} effectively modeled context dependencies and CASS-NAT~\cite{fan2020cass} generated token-level acoustic embeddings through CTC alignments, which were also confronting the redundant computation problem caused by a longer sequence.

Recently, a novel soft and monotonic alignment mechanism was proposed for sequence transduction, named continuous integrate-and-fire (CIF)~\cite{dong2020cif}. By accumulating the weight of the vector representation in each encoder step, CIF effectively locates the acoustic boundaries and extracts the acoustic information corresponding to each target token. Although it is more accurate in estimating the length of the target sequence, there still exists a deviation in the prediction of the acoustics boundaries. In this paper, in order to improve the accuracy of acoustic boundary prediction (e.g, character boundaries in Mandarin) for the CIF based models, we propose a novel auxiliary objective function to constrain the acoustic boundary of each label by using the spike information and alignment states of CTC. Morever, a recently proposed Conformer architecture~\cite{gulati2020conformer} is adopted to enhance the speech representation learning, which has better local and global modeling capabilities than Transformer. Finally, we integrate a new contextual decoder to model the linguistic and contextual dependencies within the target sequences. Evaluating on the open-source Mandarin corpora AISHELL-1, our proposed method achieves a comparable character error rates (CERs) of 4.9\% with only 1/24 latency compared with a state-of-the-art AR Conformer model. We also conduct experiments on a large scale 7500 hours internal Mandarin corpus and our model shows similar trend on in-domain test set.

The rest of this paper is organized as follows. Section \ref{sec2} introduces non-autoregressive continuous integrate-and-fire (CIF) model. Section \ref{sec3} describes our proposed method. Section \ref{sec:exp} presents our experimental setup and results. The conclusions and future work will be given in Section \ref{sec5}.

\section{Non-autoregressive Continuous Integrate-and-Fire (CIF)}\label{sec2}

Give an input speech sequence $\mathbf{x} = [\mathbf{x}_1, ..., \mathbf{x}_T]$, where $T$ means the number of frames, conventional AR models iteratively produce output tokens $\mathbf{y} = [y_1, ...,y_L]$ as:
\begin{equation}\label{eq:ar}
P_{\text{AR}}(\mathbf{y}|\mathbf{x}) = \prod_{l=1}^{L}P(y_l|y_{<l},\textbf{x}),
\end{equation}
where $L$ refers to the length of target sequence. With the previously generated tokens $y_{<l}$, AR models estimate the output token $y_l$ one-by-one, which makes it hard to compute in parallel and results in a large latency during the inference. By contrast, the NAR models aim to get rid of such temporal dependency and perform parallel computation. During the inference, NAR models will generate the probability distribution of $\mathbf{y}$ within a constant number of iterations that is not constrict to the sequence length. Mathematically, a NAR model can be defined as: 
\begin{equation}\label{eq:non-ar}
P_{\text{NAR}}(\mathbf{y}|\mathbf{x}) = \prod_{l=1}^{L}P(y_l|\textbf{x}).
\end{equation}

\begin{figure}[t]
	\centering
	\includegraphics[scale=0.51]{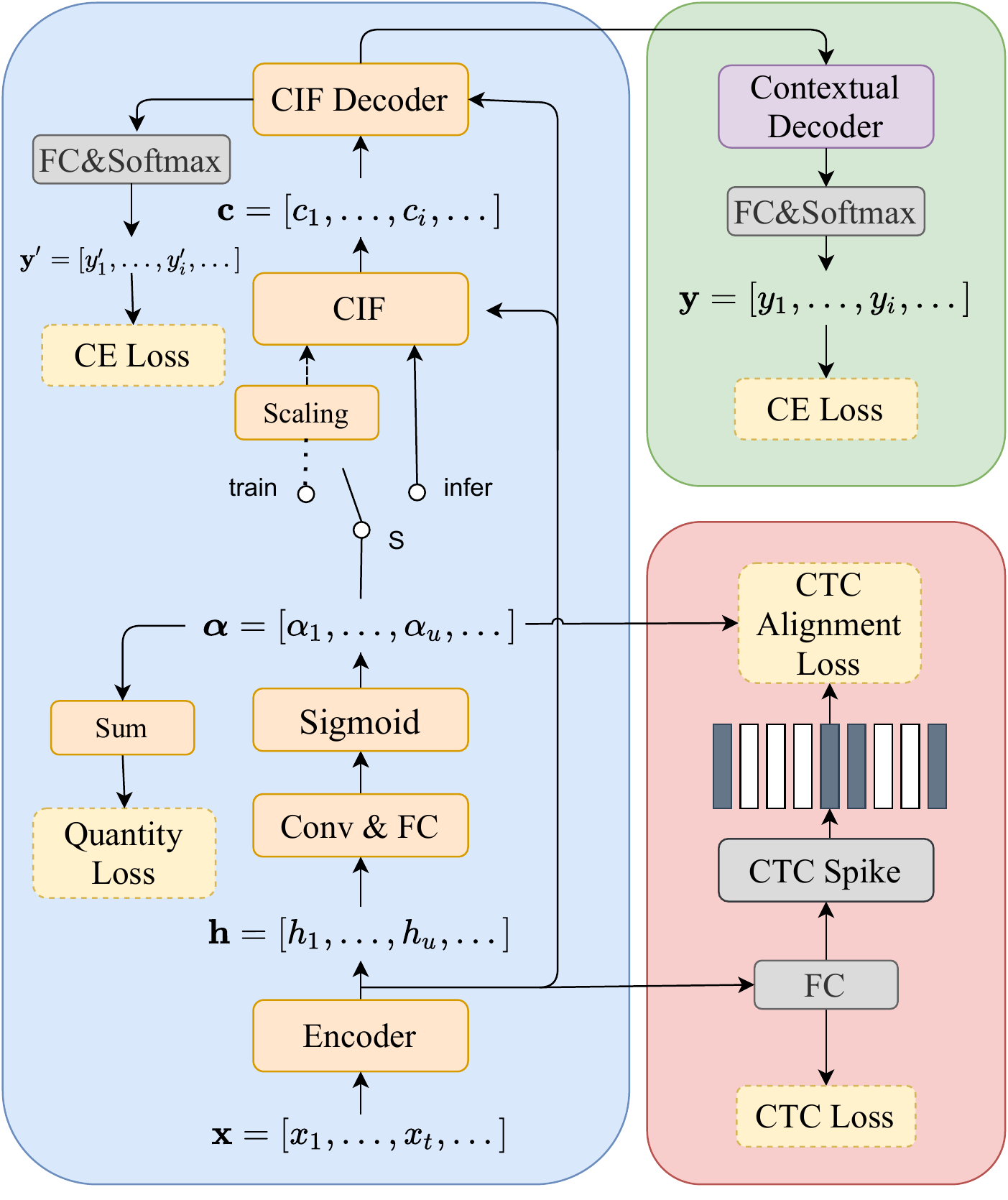}
	\caption{
		The architecture of our proposed CIF-based non-autoregressive model for the ASR task. The left part illustrates the structure of CIF module, and the right part shows the contextual decoder and CTC alignment loss.
	}
	\label{cif_structure}
\end{figure}

As a NAR model, CIF~\cite{dong2020cif} utilizes a soft and monotonic alignment mechanism between the encoder and decoder, and the encoder outputs, replacing traditional token embeddings, are directly used as the input of decoder to achieve parallel NAR computation. The left part of Fig.~\ref{cif_structure} shows a detailed structure of the CIF model. At each encoder step, CIF receives the current encoder output $h_u$ and a corresponding weight $\alpha_u$, where $u \in U$ means the index of encoder step. The weight $\alpha_u$ scales the amount of acoustic information and is accumulated until reaching a threshold $\beta=1$, which means an acoustic boundary of a specific target token. At this boundary point, the $h_u$ will be divided into two parts: one for completing the integration of current token, and the other is used for the accumulation of acoustic information for the next token. Then, the weighted summation of encoder outputs is fired as the input of decoder. Therefore, CIF is able to provide a soft alignment between acoustic frames and target labels.

In order to improve the accuracy of sequence length prediction by CIF, a quantity loss is also presented, forcing the model to predict the quantity of integrated embeddings closer to the quantity of targeted label sequence. Quantity loss $\mathcal L_{\text{Qua}}$ can be defined as:
\begin{equation}\label{eq:quantity}
\mathcal L_{\text{Qua}} = \vert  \sum_{u=1}^{U} \alpha_{u}-\tilde{S}\vert,
\end{equation}
where $\tilde{S}$ is the ground-truth length of the target sequence $\mathbf{y}$. 

\section{Proposed Method}\label{sec3}

Our proposed approach is based on the CIF model with substantial improvements to obtain more accurate acoustic boundaries and explicitly capture the linguistic contexts. Specifically, we use a CTC module to guide the learning of acoustic boundaries and introduce a novel context decoder behind the CIF decoder to model the token relationship, as shown in Fig.~\ref{cif_structure}. Besides, since Conformer encoder is proposed to learn both local and global acoustic correlations synchronously, we also adopt it to extract better acoustic hidden representations. Moreover, unlike the original CIF which only uses acoustic embedding of CIF module as in Eq.~(\ref{eq:CER-COMPUTE-cif}), we fire both the acoustic embedding $\textbf{c}$ and encoder output $\textbf{h}$ to the decoder to predict the probability of output tokens $\textbf{y} = [y_1, ...,y_i,...]$ in parallel as in Eq.~(\ref{eq:CER-COMPUTE-our}). We notice that this can lead to a further improvement of performance.
\vspace{-5pt}
\begin{align}
    P(\textbf{y}|\textbf{c}) &= \text{Decoder}(\textbf{c}) \label{eq:CER-COMPUTE-cif}, \\
    P(\textbf{y}|\textbf{c},\textbf{h}) &= \text{Decoder}(\textbf{c},\textbf{h}) \label{eq:CER-COMPUTE-our}.
\end{align}
\subsection{Conformer Encoder\label{sec31}}

Our encoder block follows the same Conformer architecture as in~\cite{gulati2020conformer, guo2021recent}, which includes a multihead self-attention (MHSA) module, a convolution (Conv) module, and a pair of positionwise feed-forward (FFN) module in the Macaron-Net style. Meanwhile the MHSA module can learn long-range global context, the Conv module is good at extracting fine-grained local features. By adopting Conformer, we expect to improve the prediction performance of CTC and CIF modules as well. In addition, incorporating the relatively positional embeddings to MHSA modules further improves the generalization of the model on variable lengths of input sequences.

\subsection{CTC alignment loss\label{sec32}}

\begin{figure}[t]
	\centering
	\includegraphics[scale=0.55]{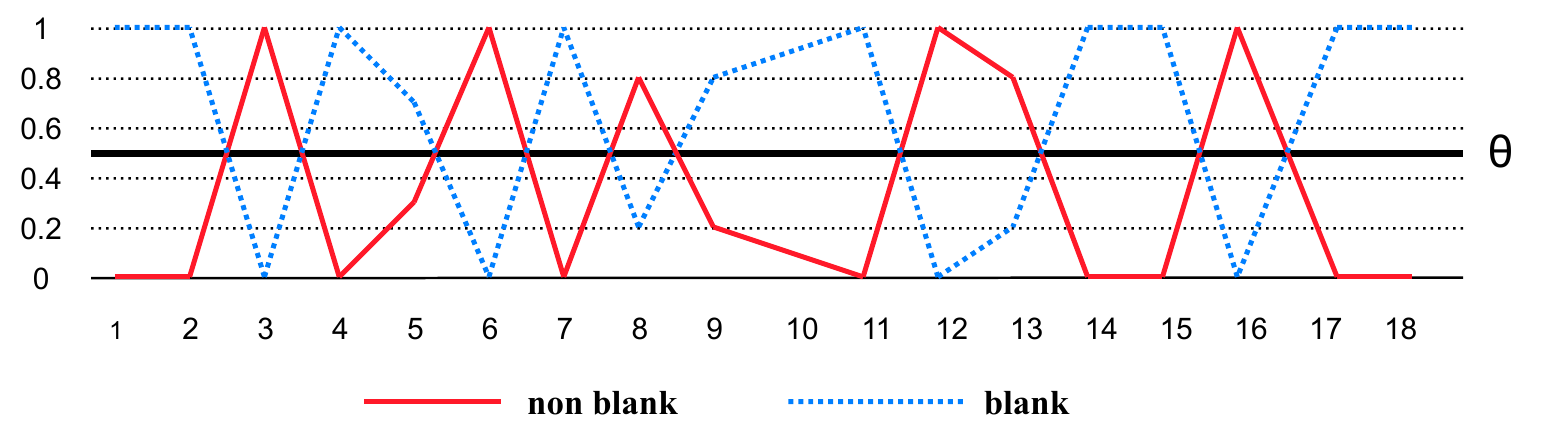}
	\vspace{-15pt}
	\caption{
		The spike-like trigger probability curve. The red curve represents the non-blank label probability, and the solid black line represents the trigger threshold $\theta$. The probability of a non-blank token greater than $\theta$ is a spike.
	}
	\label{CTC_Spike}
\vspace{-15pt}
\end{figure}

Due to the speaker’s speaking rate, accent, silence and noise, which often happen in real-world applications, the alignment learning strategy of CIF may result in an inaccurate acoustic boundary estimation, which severely influences the convergence speed. To figure out these problems and improve the performance, we present an additional CTC alignment loss function (as shown in Fig~\ref{cif_structure}) to guide the CIF-based model to predict the acoustic boundary closer to the actual boundary by making full use of the CTC spike information. Since the CTC spike is usually within the acoustic boundary of the label, we can roughly determine the boundary of a label by two consecutive CTC spikes. The CTC spike module is shown in Fig.~\ref{CTC_Spike}. Given a CTC spike sequence $P_{\text{s}}=[0,0,1,0,0,1,0,1,0,...]$, where 1 means the probability of non-blank token is greater than a specific threshold $\theta$, we constrain the CIF weights $\bm{\alpha}$ by the label boundary sequence $P_{\text{b}}=[0,2,5,7,...]$\footnote{The first element of $P_\text{b}$ is set to 0 for identifying the beginning of the spike sequence.}, where each number is the index of non-blank token. The additional alignment loss $\mathcal L_{\text{Ali}}$ can be formulated as:

\begin{align}
\mathcal  L_{\text{Ali}} = \sum_{i=0}^{L}\vert \sum_{j=P_{\text{b}}(i)}^{P_{\text{b}}(i+1)} \alpha (j) -1 \vert.
\end{align}
where $L$ is the ground-truth length of the targets sequence $\textbf{y}$. By providing the alignment label boundary constraints, we combine the boundary information from both CTC spikes and CIF to promote the model to locate boundaries in a parameter-efficient way.

\vspace{-5pt}

\subsection{Contextual decoder\label{sec33}}

To further help the CIF decoder to capture the token relationship, we propose a contextual decoder after CIF decoder to capture the contextual relationship within a sequence. Although the outputs of the CIF decoder can be regarded as an integrated representation of acoustic and linguistic information corresponding to the tokens, the context correlation between tokens is weak. The contextual decoder leverages a stack of self-attention blocks, and its input of the query, key, and value is the high-level representations from the CIF decoder. Meanwhile, contextual decoder does not require the acoustic output representation of encoder since it already contains enough acoustic information of each token. Moreover, considering the deep layers of the model, we calculate the cross-entropy (CE) loss of between the outputs of CIF decoder and contextual-decoder to assist the backward update of the gradient and speed up the convergence. The cross-entropy loss can be formulated as $\mathcal L_{\text{CE}}=\mathcal L_{\text{CE}}(\bm{y^{'}}) + \mathcal L_{\text{CE}}(\bm{y^{''}})$, where $\bm{y^{'}}$ and $\bm{y^{''}}$ indicate CIF decoder output and contextual decoder output respectively. Note that our model with contextual decoder needs to train more epochs or use a pre-trained encoder to speed up the model convergence.

\subsection{Training strategy}




As described in Section~\ref{sec2}, considering the accuracy of sequence length prediction for CIF module and the lack of left-to-right constraints for the attention model, our model adopts the quantity loss and CTC loss. In addition, we introduce a CTC alignment loss $\mathcal L_{\text{ali}}$ to further promote the learning of acoustic boundary as described in Section~\ref{sec32}. Finally, we also compute the CE loss $\mathcal L_{\text{CE}}$ for the CIF decoder and the contextual decoder to boost the model learning. Therefore, our model is trained with a combination of four different losses:
\vspace{-2pt}
\begin{equation}\label{eq:CER-COMPUTE}
\begin{aligned}
\mathcal L =\mathcal L_{\text{CE}} + {\lambda}_{1} \mathcal L_{\text{Ali}} +  {\lambda}_{2} \mathcal L_{\text{CTC}} + {\lambda}_{3}\mathcal L_{\text{Qua}},
\end{aligned}
\end{equation}
where ${\lambda}_{1}$, ${\lambda}_{2}$ and ${\lambda}_{3}$ are interpolation factors. In this study, these hyper-parameters are set to $1$ and the effects of these auxiliary functions will be investigated in the following experiments. 

\vspace{-4pt}
\section{Experiments}\label{sec:exp}

\subsection{Dataset}\label{ssec:data}
In this paper, we evaluate our Conformer-CIF based model on two Mandarin speech recognition corpora: an open-source AISHELL-1 corpus~\cite{bu2017aishell} and a large-scale internal 7500 hours corpus~\cite{wang2021cascade}. The AISHELL-1 corpus contains 150 hours of speech recorded by 340 speakers as training set, 18 hours of speech recorded by 40 speakers as development set, and about 10 hours speech as test set. The speakers of different sets are not overlapped and each set is recorded in a relatively quiet environment. The 7500 hours internal corpus contains reading speech data in various domains, such as entertainment, journalism, literature, technology, free conversation, etc. For experiments on the 7500 hours corpus, we report character error rate results (CERs) on four test sets, namely AISHELL-1 test set (TA1)~\cite{bu2017aishell}, AISHELL-2 test set (TA2)~\cite{du2018aishell}, an internal voice input test set (VI), and a voice assistant (VA) test set. The VI test set consists of about 3.4 hours data with 3063 sentences covering lots of proper nouns and named entities, which is used to verify the language generalization ability of the model. The VA test set is collected under various noisy scenarios, containing about 3.9 hours data with 5000 voice assistant commands.

\subsection{Experimental Setup}\label{ssec:setup}
In our work, the 80-dimensional log Mel-filter bank feature (Fbank) plus 3-dimensional pith feature are used as the input feature. The window size is 25 ms with a shift of 10 ms. We use 4231 characters extracted from the training transcriptions as the modeling units. Our CIF-based NAR model is comprised of 12 encoder layers, 6 CIF decoder layers, and 6 contextual decoder layers. Particularly, when performing experiments on the 7500 hours corpus, the number of encoder layers and contextual decoder layers is set to 20 and 4, respectively. The common parameters of the encoder and decoder layers are: $d^{head} = 4, d^{attn} = 256, d^{ff} = 2048$ for the number of attention heads, dimension of attention module, and dimension of feedforward layer, respectively. We use the Adam optimizer~\cite{kingma2015adam} and the warmup scheduler~\cite{vaswani2017attention} to train the model for 80 epochs. The warmup step is set to 25k iterations. SpecAugment~\cite{park2019specaugment} and speed perturbation~\cite{ko2015audio} with the factor of 0.0, 1.0, and 1.1 are also applied to avoid over-fitting. For a more fair comparison, we re-implement the CTC and Spike-Triggered~\cite{tian2020spike} models with the same parameters and Conformer structure as used in our model. All the experiments are conducted using the open-source ESPnet~\cite{watanabe2018espnet} toolkit on a single NVIDIA RTX 2080Ti GPU.

\begin{table}[t]
\caption{The character error rates (CERs) of the systems on AISHELL-1. Real-time factor (RTF) is computed as the ratio of the total inference time to the total duration of the test set.}
\vspace{0.4cm}
\begin{threeparttable}[t]
\begin{tabular}{lccc}
\hline
\multicolumn{1}{c}{\multirow{2}{*}{Model}} & \multicolumn{2}{c}{CER (\%)}                       & \multirow{2}{*}{RTF} \\ \cline{2-3}
\multicolumn{1}{c}{}                       & Dev                     & Test                    &                      \\ \hline
\textit{Autoregressive}                    & \multicolumn{1}{l}{}    & \multicolumn{1}{l}{}    & \multicolumn{1}{l}{} \\
\quad Kaldi chain~\cite{bu2017aishell}                                & -                       & 7.6                     & -                    \\ 
\quad Conformer                                  & 4.8                     & 5.1                     &  0.2768                    \\
\quad \quad + Contextual decoder                           & 4.4                     & 4.8                     &  0.4306                           \\ \hline
\textit{Non-autoregressive}                &                         &                         &                      \\
\quad Transformer-A-FMLM~\cite{chan2016listen}                         &  6.2                      &  6.7                      &  -                   \\
\quad Transformer-Insertion~\cite{fujita2020insertion}                         &  6.1                      &  6.7                      &  -                   \\
\quad Transformer-LASO~\cite{bai2020listen}                                       & 5.8                     & 6.4                     &  -                    \\
\quad Transformer-CASS-NAT~\cite{fan2020cass}                                       & 5.3                     & 5.8                     &  -                    \\
\quad Conformer-CTC                                        & 5.1                     & 5.7                     &  0.0125                    \\
\quad Conformer-Spike-Triggered~\cite{tian2020spike}\tnote{$\dagger$}                  &5.0 & 5.6 & 0.0152
\\ \hline
\textit{Non-autoregressive (proposed)}                &                         &                         &                      \\
\quad Conformer-CIF                         & 4.8                     & 5.3                    &  0.0166                    \\
\quad \quad + Contextual decoder                            & 4.6                     & 5.0                       &  0.0181                    \\
\quad \quad \quad + CTC alignment loss                         & \textbf{4.5}                     & \textbf{4.9}                     & 0.0182                     \\ \hline
\end{tabular}
\begin{tablenotes}
    \footnotesize
		\item $\dagger$: This models is re-implemented by ourselves with the same parameter structure as our model.
\end{tablenotes}
\end{threeparttable}
\label{tab:result_cif}
\end{table}

\subsection{Results on AISHELL-1}\label{ssec:aishell}


\begin{figure*}[t]
	\centering
	\includegraphics[width=1.0\linewidth]{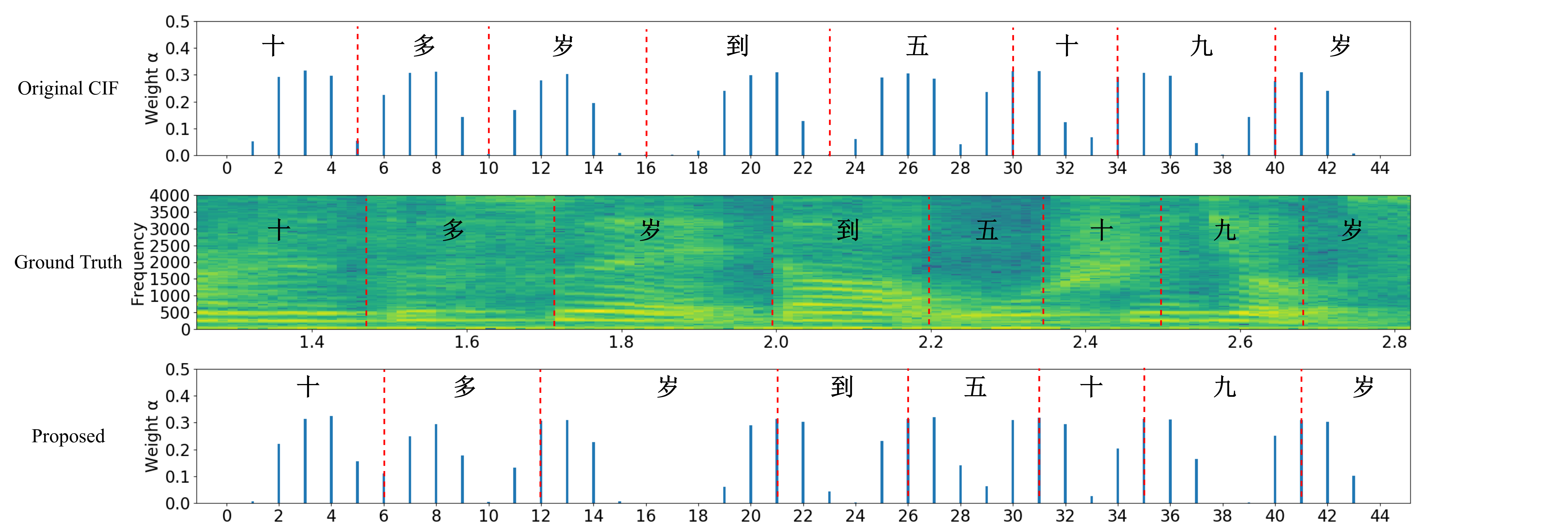}
	\caption{
	Visualization on the prediction of weight $\alpha$ for each encoded state for the original CIF model (upper) and our CIF model (lower) with contextual decoder and CTC alignment loss. These figures are drawn for the utterance index BAC009S0764W0356 in the AISHELL-1 test set. We find that the original CIF model (upper), as mentioned in \cite{dong2020cif}, is more prone to be located ahead of time, but our CIF model (lower) predicts the boundary of each token more accurately. Red dotted vertical line: acoustic boundary of the token.
	}
	\label{cif_alignment}
\end{figure*}

To explore the efficiency of our proposed Conformer-CIF based model, we first evaluate it on the open-source AISHELL-1 corpus. As show in Table~\ref{tab:result_cif}, we can find that our model outperforms the hybrid Kaldi chain system and even reaches the state-of-the-art AR Conformer model, achieving CERs of 4.8\% and 5.3\% on dev and test sets, respectively. Comparing with other NAR models, our Conformer-CIF based model is still better than other methods with a slight increment of RTF. When incorporating with the contextual decoder and CTC alignment loss, we can obtain further improvement, achieving CERs of 4.5\% and 4.9\% with about $24\times$ faster in inference speech.

\subsection{Results on 7500 hours}\label{ssec:7500}
We further evaluate the proposed model on a large-scale internal 7500 hours corpus, as shown in Table~\ref{tab:result_bigdata}. In addition to the clean test sets TA1 and TA2, we also involve two more challenging test sets VI and VA. From the table, we can see a quite similar trend as experiments on the AISHLLE-1 corpus. The proposed Conformer-CIF based model outperforms the cascade RNN-T model on 3 test sets and even shows good potential on the most challenging sets VI and VA, where VA is corrupted by real-world noise or reverberation. Comparing with the state-of-the-art AR Conformer model, our model achieves competitive results on TA1 and TA2 tests with only 1/18 latency. Since VI and VA are collected under more complicated acoustic conditions from a mismatched domain, comparing with AR model, the performance of the NAR model is reduced due to lack of context information within the sequences. Morever, by comparing the results with Conformer-CTC NAR models, our model yields up to 9.5\%, 8.0\%, 5.8\%, and 1.5\% relative CERs reductions on four test sets, respectively.
\begin{table}[t]
\caption{The character error rates (CERs) of the systems on the the 7500-hour corpus. Real-time factor (RTF) is computed as the ratio of the total inference time to the total duration of the test set.}
\vspace{0.4cm}
\resizebox{1.0\linewidth}{!}{
    \begin{tabular}{lccccl}
    \hline
    \multicolumn{1}{c}{\multirow{2}{*}{Model}} & \multicolumn{4}{c}{CER(\%)} & \multirow{2}{*}{Param.(M)/RTF} \\ \cline{2-5}
    \multicolumn{1}{c}{}                       & TA1   & TA2  & VI   & VA    &                                \\ \hline
    \textit{Autoregressive}                &                         &                         &                      \\
    \quad Cascade RNN-T~\cite{wang2021cascade}                              & 4.6  & 9.2 & 8.7 & 28.1 & 95.5M / \quad -                     \\
    \quad Conformer                                  & 5.1  & 8.4 & 6.7 & 25.7 & 71.2M / 0.4487                     \\
    \hline
    \textit{Non-autoregressive}                &                         &                         &                      \\
    \quad Conformer-CTC                                  & 6.1  & 9.7 & 8.6 & 27.6 & 55.8M / 0.0201                     \\
    \quad Conformer-CIF                        & \textbf{5.5}  & \textbf{8.9} & \textbf{8.1} & \textbf{27.2}  & 73.6M / 0.0253                    \\ \hline
    \end{tabular}
}
\vspace{-0.4cm}
\label{tab:result_bigdata}
\end{table}

\subsection{Result Analysis}\label{ssec:analysis}
The results on Section~\ref{ssec:aishell} and Section~\ref{ssec:7500} have demonstrated that the proposed method significantly improves the performance with a negligible RTF increment. In this Section, we will investigate the effectiveness of the CTC alignment loss and the contextual decoder.

Fig.~\ref{cif_alignment} shows a visualization of the encoded state weight $\alpha$ for the original CIF model and our proposed Conformer-CIF model with contextual decoder and CTC alignment loss. For the upper and lower parts, the red dash lines are acoustic boundaries estimated by the models, which means the accumulated weight reaches 1.0 at this point. The middle part is the spectrogram of the audio and the acoustic boundaries are marked manually. We can see that the original CIF model is prone to be located ahead of time, as mentioned in \cite{dong2020cif}, while our model gives a more accurate estimation of the acoustic boundaries. By using the CTC spike information to guide the learning of acoustic boundaries, our model can not only converge faster, but also predict the length of the target sequence correctly, which will dramatically reduce the insertion and deletion errors.

\begin{figure}[t]
	\centering
\vspace{-7pt}
	\includegraphics[scale=0.88]{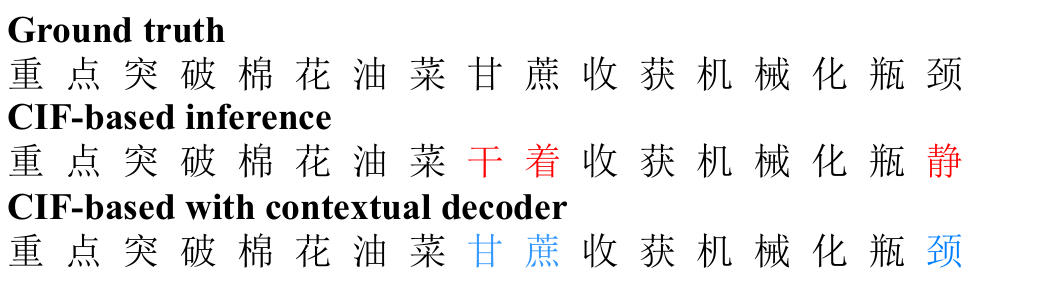}
		\vspace{-15pt}
	\caption{
		Decoding example for BAC009S0903W0239 in the AISHELL-1 test set. Red indicates characters with errors and blue indicates ones recovered with a contextual decoder.
	}
	\label{decoder_example}
\vspace{-0.2cm}
\end{figure}

Moreover, we also notice that most of the inference errors are substitution errors with similar pronunciation, as shown in Fig~\ref{decoder_example}. The upper part shows the ground truth sequence, while the middle and lower parts are the output hypotheses obtained by different models. Considering the lack of context dependencies within a sentence, we adopt an additional contextual decoder to improve the linguistic and contextual relationship and aim to recover such substitution errors. From figure~\ref{decoder_example}, we can see that the proposed contextual decoder can successfully eliminate substitution errors.

\section{Conclusions}\label{sec5}

In this paper, we propose a boundary and context aware training approach for CIF based non-autoregressive models to improve the accuracy of the acoustic boundary prediction. Our Conformer-CIF based model utilizes the CTC spike information to guide the learning of acoustic boundaries in the CIF, and integrates a new contextual decoder to model the linguistic and contextual dependencies within the target sequence. Meanwhile, we also adopt a recently proposed Conformer architecture to improve the capacity of acoustic modeling. Evaluating on the open-source Mandarin corpora AISHELL-1 and an internal 7500 hours Mandarin corpus, our proposed approach achieves comparable character error rates with only 1/24 latency compared with a state-of-the-art autoregressive Conformer model. In the future, we will integrate the external language model into our proposed model to improve the language generalization ability.

\section{Acknowledgement}

This work was supported by MoE-CMCC “Artificial Intelligence” Project (MCM20190701).


\bibliographystyle{IEEEbib}
\bibliography{strings,refs}

\end{document}